\documentclass[11pt]{article}
\usepackage{amsmath,psfig}

\bigskip

\begin{document}
\title{Electrically Charged Einstein-Born-Infeld Black Holes with Massive Dilaton}

\author{Stoytcho S.~Yazadjiev\thanks{E-mail: yazad@phys.uni-sofia.bg} ,
Plamen P.~Fiziev\thanks{E-mail: fiziev@phys.uni-sofia.bg}\\
{\small\textit {Department of Theoretical Physics,
                Faculty of Physics,}}\\ [-1.mm] {\small\textit {Sofia University, Bulgaria}}
                \\\\[-3.mm]
 Todor L.~Boyadjiev${^1}$\thanks{E-mail: todorlb@fmi.uni-sofia.bg} ,
 Michail D.~Todorov${^{2}}$\thanks{E-mail: mtod@vmei.acad.bg} \\
{\small\textit {${^1}$Faculty of Mathematics and Computer Science,
Sofia University, Bulgaria }}\\ [-1.mm]{\small\textit
{${^2}$Faculty of Applied Mathematics and Computer Science,}}\\
[-1.mm] {\small\textit {Technical University, Sofia, Bulgaria }}}

\date{}

\maketitle

\begin{abstract}
We numerically construct static and spherically symmetric
electrically charged black hole solutions in Einstein-Born-Infeld
gravity with massive dilaton. The numerical solutions show that
the dilaton potential allows many more black hole causal
structures than the massless dilaton. We find  that depending on
the black hole mass and charge and the dilaton mass  the black
holes can have either one, two, or three horizons. The extremal
solutions are also found out. As an interesting peculiarity we
note that there are  extremal black holes with an inner horizon
and with  triply degenerated horizon.
 \end{abstract}

\bigskip

\section{Introduction}

In recent years there has been an increasing interest in the
Born-Infeld (BI) type of generalizations of Abelian and
non-Abelian gauge theories. Such generalizations appear naturally
in the context of the (super)string theory. The BI action
including a dilaton and an axion field, appear in the couplings
of an open superstring and an Abelian gauge field \cite{FT}.
This  action, describing a BI-dilaton-axion system coupled to
Einstein gravity, can be considered as a non-linear extension in
the Abelian field of Einstein-Maxwell-dilaton-axion (EMDA)
gravity. It has also been shown that the world volume action of a
D-brane is describe by a kind of BI action arising as a sum over
disc string diagrams \cite{P}.

Without any doubt the black holes are one of the most interesting
objects in both pure Einstein-Born-Infeld (EBI) and
Einstein-Born-Infeld-dilaton(-axion) (EBID(A)) gravity. Black
holes in pure EBI theory were studied earlier in
\cite{D}-\cite{O}. The motivation for considering EBID(A) black
holes is to understand the possible causal structures in the
presence of a dilaton (and axion) coupled to a nonlinear
electromagnetic field. The nonlinearity of the electromagnetic
field may lead to surprising phenomena. It was shown that a kind
of nonlinear electromagnetism produces nonsingular black holes
satisfying the weak energy condition \cite{ABG1} - \cite{ABG3}. It
should be noted that this may contradict the strong version of the
cosmic censorship conjecture.

Recently EBID  black holes in four dimensions  have been
considered in \cite{TT1}, \cite{TT2}, and \cite{CG}. The three
dimensional EBID black holes  are studied in \cite{YI}. It has
been shown that the four-dimensional EBID black holes have
remarkable properties which are not observed for the
corresponding EMD and EBI solutions. By examining the internal
structure of the black holes, it has been found that there is no
inner horizon and that the global structure is the same as the
Schwarzschild one in both electrically and magnetically charged
case.

In the black hole investigations performed so far the dilaton has
been considered as massless. The massless dilaton, however,
contradicts the experiments. That is why, it is the EBI black
holes coupled to a massive dilaton, which are the most interesting
from a physical point of view.

The purpose of the present letter is to present  numerical EBI
black hole solutions  with a massive dilaton and to show that the
massive dilaton allows many more black hole causal structures than
the massless dilaton. We find that depending on the black hole
mass and charge and the dilaton mass the EBI black holes can have
either one, two or three horizons. The extremal black hole
solutions were also found out.

It should be noted that black holes coupled to a massive dilaton
but within the framework of ordinary EMD gravity were studied
earlier in \cite{GH} and \cite{HH}. The authors of \cite{GH} using
a qualitative analysis argue that there may exist EMD black holes
with three horizons and black holes with a triply degenerated
horizon in presence of a massive dilaton. The authors of \cite{HH}
present numerical EMD black hole solution with one and two
horizons and also argue that  space-times with three horizons
cannot be ruled out. Note, however, that no black solutions with
three horizons or with a triply degenerated horizon were presented
in \cite{GH} and \cite{HH}. We hope that the presented (numerical)
black holes solutions (within the framework of EBID gravity) in
this letter  are the first explicit examples of black holes with
three horizons and triply degenerated horizon in presence of a
massive dilaton. This way we also confirm, though within the
framework of EBID gravity, that the massive dilaton, in general,
allows black holes with three horizons as well as with a triply
degenerated horizon as it is argued in \cite{GH} and \cite{HH}.

\section{Basic equations}
We consider the following EBID action
\begin{equation}
\label {ACT} S = \frac{1}{16 \pi}\int d^4x \sqrt{-g}\, [R -
2\nabla^{\mu}\varphi \nabla_{\mu}\varphi  - U(\varphi) + {\cal
L}_{BI}],
\end{equation}
where $R$ is Ricci scalar curvature with respect to the space-time
metric $g_{\mu\nu}$, $\varphi$ is the dilaton field and
$U(\varphi)$ is the dilaton potential. The BI part of the action
is given by (see \cite{TT1} and references therein):
\begin{equation}
{\cal L}_{BI} = 4b e^{2\alpha\varphi} \left\{1  -  \left[1 +
\frac{e^{-4\alpha\varphi}}{2b}F^2  - \frac{e^{-8\alpha\varphi}}
{16b^2}(F\star F)^2  \right]^\frac{1}{2} \right\}.
\end{equation}
Here $\star F$ is the dual to the Maxwell tensor and  $\alpha$ is
the dilaton coupling parameter. In the context of the string
theory, the BI parameter $b$ is related to the string tension
$\alpha^\prime$ by $b= 1/2\pi\alpha^\prime$. Note that in the
$b\to \infty$ limit our action reduces to the EMD system with
massive dilaton. It should be pointed that EBID action does not
posses an electric-magnetic duality. That is why one may expect
that the electrically and magnetically charged black holes will be
different.

We are interested in static and spherically symmetric
configurations and therefore assume the general static,
spherically symmetric parameterization of the space-time metric
\begin{equation}
\label{METRIC} ds^2 = - f e^{-2\delta} dt^2 + f^{-1}d\rho^2 +
\rho^2 d\Omega,
\end{equation}
\noindent where $f$ and $\delta$ depend on $\rho$ only. As
boundary conditions at infinity, we require the metric be
asymptotically flat and the dilaton vanish, i.e.,
$\varphi(\infty)= 0$. Note that the condition $\varphi(\infty)=
0$ is consistent with the asymptotic flatness when and only when
the minimum of the dilaton potential is chosen to be $\varphi=0$.

In the present letter we consider only electrically charged
configurations. In this case the gauge potential has the form
\begin{equation}
A = -\, \Phi(\rho) dt
\end{equation}
where $\Phi$ is the electric potential. From the BI equations we
obtain
\begin{equation}
\label {ELF} {d\Phi\over d\rho} = -  {Q_{e}\over \sqrt{\rho^4 +
{Q_{e}^2\over b}} }\,e^{2\alpha\varphi}e^{-\delta}.
\end{equation}
The electric charge $Q_{e}$ is defined by $Q_{e}= - \lim_{\rho \to
\infty} \rho^2 {d\Phi(\rho)\over d\rho}$.

Using metric (\ref{METRIC}) and the expression for the electric
field (\ref{ELF}) in the field equations determined by action
(\ref{ACT}) we get the equations for the structure of a
spherically symmetric black hole. Before we explicitly write them,
we are going to introduce a dimensionless radial coordinate $$r=
\sqrt{b}\rho$$ and a dimensionless dilaton potential given by
$$U(\varphi)= 2m^2_{D}V(\varphi)$$ where $m_{D}$ is the dilaton mass.
From now on, the prime will denote a differentiation with respect
to the dimensionless coordinate $r$. We also define  other
dimensionless quantities by
\begin{eqnarray}
q_{e}= \sqrt{b}Q_{e}\,\,\, ,\,\,\, \gamma= {m_{D}\over \sqrt{b}}
\,.
\end{eqnarray}

It is also convenient to introduce the local mass $m(r)$ defined
by $f=1-2m(r)/ r$.

With all these definitions the equations for  a spherically
symmetric black hole are reduced to the following
\begin{eqnarray}
m^{\prime}(r) \!\!&=&\!\! \tfrac{1}{2} f r^2 (\varphi^\prime)^2 +
\tfrac{1}{2}\gamma^2 V(\varphi)r^2 + e^{2\alpha\varphi}\left(
\sqrt{r^4 + q^2_{e}} - r^2\right), \nonumber \\ \delta^\prime (r)
\!\! &=&\!\! - r (\varphi^\prime)^2, \label{GE} \\
\varphi^{\prime\prime}(r) \!\! &=&\!\! -\frac{2}{r}\varphi^\prime
+ f^{-1}\left(\frac{f-1}{r} + \gamma^2V(\varphi)r+
 2e^{2\alpha\varphi} \frac{\sqrt{r^4 + q^2_{e}}- r^2 }{ r }
 \right)\varphi^\prime \nonumber \\
&&+ f^{-1}\left(\frac{\gamma^2}{2} \frac{d V(\varphi)}{d\varphi} +
2\alpha e^{2\alpha\varphi}\frac{\sqrt{r^4 + q^2_{e}}- r^2}{r
}\nonumber \right) \,.
\end{eqnarray}

In order to satisfy the asymptotic flatness  we impose the
following boundary conditions at the spatial infinity:
\begin{equation}
m(\infty) = M, \,\,\,\,\,\, \delta(\infty)= 0, \,\,\, \,\,\,
\varphi(\infty) = 0\, .
\end{equation}

Here $M$ is the dimensionless black hole mass which is related to
the dimensionfull mass ${\cal M}$ by  ${\cal M}= M/\sqrt{b}$.

We also assume the existence of a regular  or  degenerated event
horizon at $r = R_{h}$ and we have
\begin{equation}
f_h = 0, \quad \varphi_{h}< \infty , \quad \delta_{h} < \infty,
\quad f^{\prime}_h \varphi^{\prime}_{h} = \frac{1}{2}\gamma^2
\frac{dV}{d\varphi}(\varphi_{h}) + 2\alpha e^{2\alpha\varphi_{h}}
\frac{\sqrt{R^4_{h} + q^2_{e}} - R^2_{h}}{R^2_{h}} \,\, \, .
\end{equation}

\noindent where the variables with subscript $h$ shows that they
are evaluated at the horizon.

Under this conditions we obtain the black hole solutions
numerically using the continuous analogue of Newton method
\cite{Puz}, \cite{BTFY}. We have carefully tested our numerical
code reproducing independently the results from the article
\cite{TT1}-\cite{TT2}.

\section{Numerical results}

In what follows we consider black holes with a dilaton coupling
parameter $\alpha = -1$ and dilaton potential $V(\varphi)=
\varphi^2$.
\bigskip
\begin{figure}[htb]
\centerline{\psfig{figure=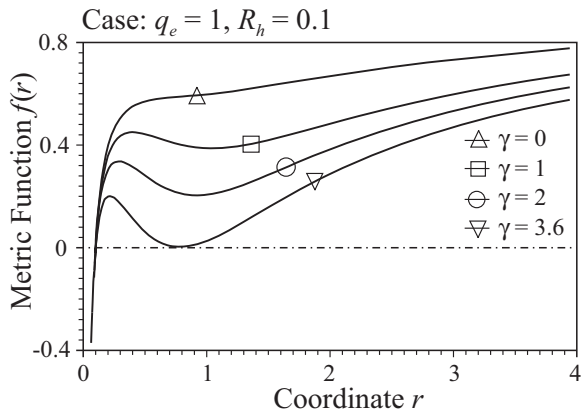,width=3in}} \caption{\small
{Radial dependence of the metric function $f$ for different values
of the parameter $\gamma$. The solution corresponding to
$\gamma=3.6$ is extremal.}} \label{fig1}
\end{figure}

In Fig.1 we show the results of the numerical integration for the
radial dependence of the metric function $f$ for  different values
of the parameter $\gamma$ and for a black hole charge $q_{e}=1$.
The black hole radius is almost the same for those values of
$\gamma$ presented in the figure. The curves $f(r)$ for
$\gamma=0$, $\gamma=1$, and $\gamma=2$ represent EBI black holes
with only one horizon and mass $M=0.6$, $M=0.76$, $M=0.86$,
respectively. The curve $f(r)$ for $\gamma=3.6$ represents an
extremal black hole solution with mass $M=0.96$. As an interesting
peculiarity we note that this extremal black hole has an inner
horizon.

\bigskip
\begin{figure}[ht]
\centerline{\psfig{figure=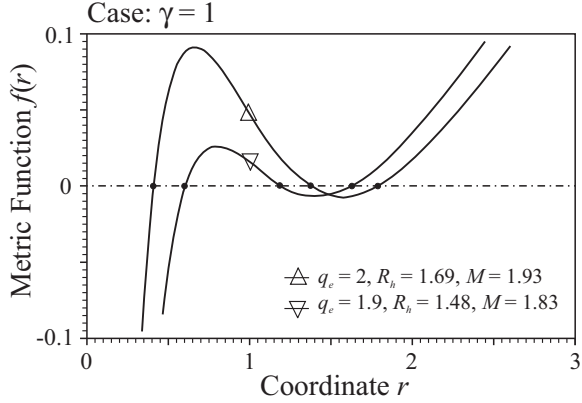,width=3in}}
\caption{\small{Radial dependence of the metric functions $f$ for
black holes with three horizons.}} \label{fig2}
\end{figure}

Solutions representing EBI black holes with three horizons are
shown in Fig.2. The parameters $q_{e}= 2$ and $R_{h}= 1.69$
correspond to a black hole with mass $M=1.93$, while the others
correspond to black hole mass $M=1.82$.

\bigskip
\begin{figure}[ht]
\centerline{\psfig{figure=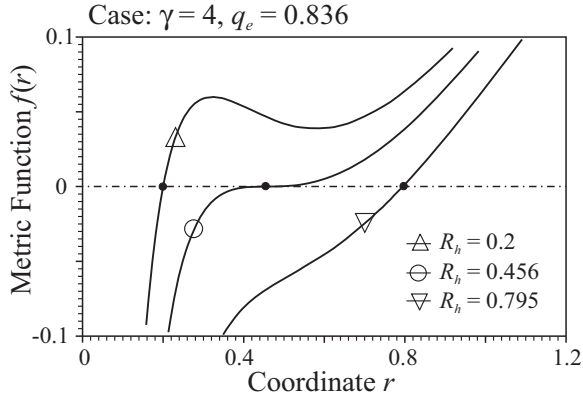,width=3in}} \caption{\small{The
curve corresponding to $R_h=0.456$ represents an extremal black
hole with a triply degenerated horizon.}} \label{fig3}
\end{figure}

In Fig.3 we show the radial dependence of the metric function $f$
for an extremal black hole with a triply degenerated horizon. The
mass of this black  hole is $M=0.79$.

\bigskip
\begin{figure}[ht]
\centerline{\psfig{figure=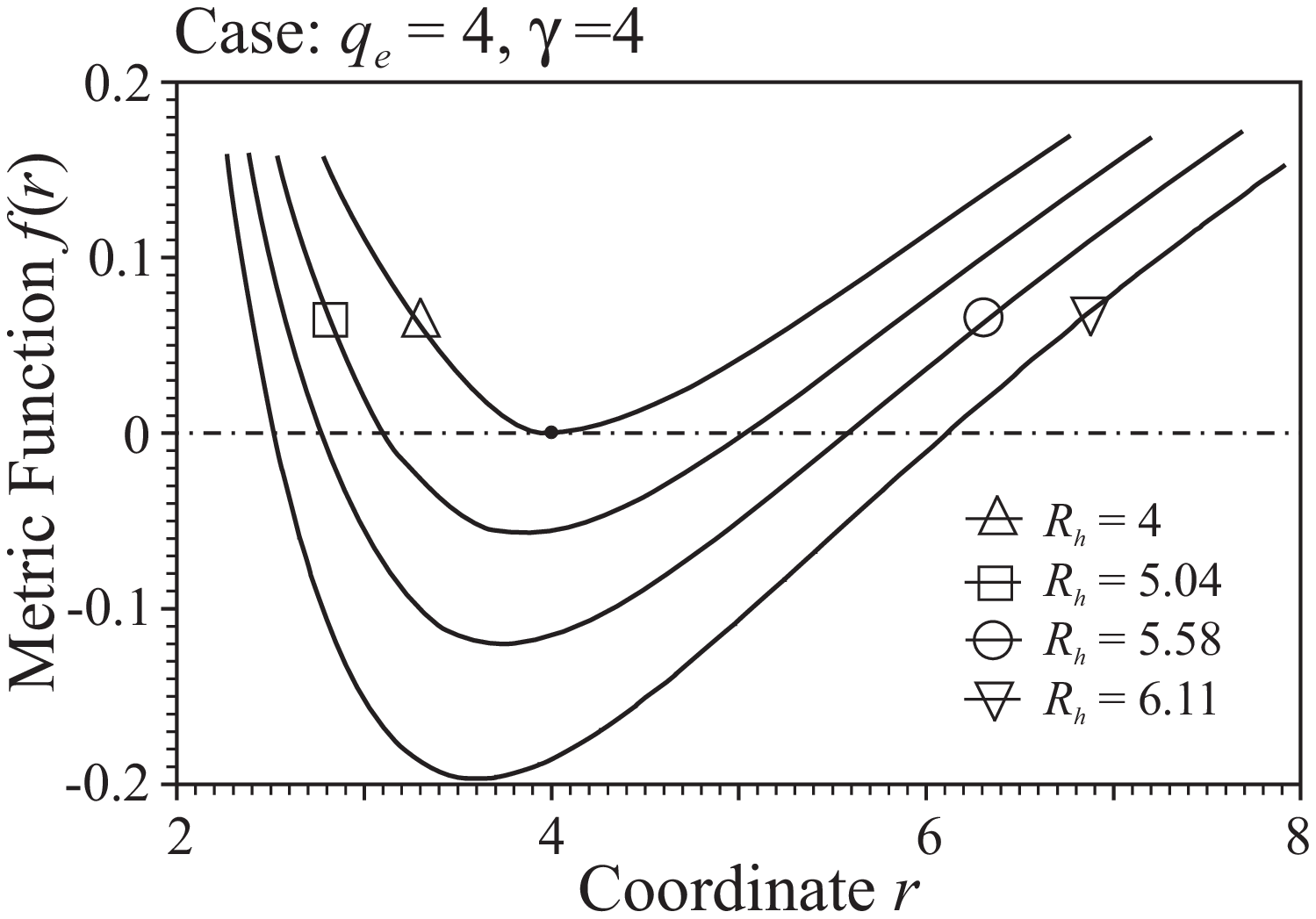,width=3in}} \caption{\small
{The function $f$ represents Reissner-Nordstr{\"o}m type black
holes. The solution with $R_{h}=4$ is extremal.}} \label{fig4}
\end{figure}

We have also found  EBI black holes with two horizons whose causal
structure is of the Reissner-Nordstr{\"o}m type. The metric
function $f(r)$ for these black holes is presented in Fig.4. The
curves with $R_{h}=5.04$, $R_{h}=5.58$, and $R_{h}=6.11$
correspond to black holes with masses $M=4.1$, $M=4.22$, and
$M=4.36$, respectively. The solution with radius $R_{h}=4$
represents an extremal black hole with $M=4$.

So far our considerations of the black holes causal structures
have been performed in Einstein frame. The Einstein frame and the
string frame are related via  the conformal factor
$e^{-2\varphi}$. Since the dilaton field and its derivatives are
regular everywhere except at the center where they  diverge, the
black hole causal structures will remain the same in the string
frame.

\section*{Conclusion}
In this letter we have presented numerical EBI black hole
solutions with a massive dilaton. These solutions show that the
massive dilaton allows many more black hole causal structures than
the massless one. Depending on the black hole mass, charge and
the dilaton mass, EBI black holes can have either one, two, or
three horizons. Extremal black hole solutions have been presented,
too. There are extremal black holes  with an inner horizon as well
as extremal black holes with a triply degenerated horizon.

An extended version of this letter containing a qualitative
analysis and detailed numerical study of the global physical
characteristics (black hole mass, charge, temperature, etc.) of
both non-extremal and extremal electrically and magnetically
charged EBI black holes and for different dilaton  potentials,
will be published elsewhere.

\medskip
\section*{Acknowledgments}

We would like to thank  R. Rashkov for the useful comments. We
would also like to thank  D. Wiltshire, G. Clement  and D.
Gal'tsov for pointing out some useful references.

\end{document}